\documentclass[letterpaper,12pt]{article}
\usepackage[ansinew]{inputenc}
\usepackage[english]{babel}
\usepackage{amsmath}
\usepackage{amsthm}
\usepackage{amssymb}
\usepackage{enumerate}
\usepackage[margin=1in]{geometry}
\usepackage{graphicx}
\usepackage{makeidx}
\usepackage{float}
\usepackage{authblk}
\usepackage{datetime}
\usepackage{url}

\restylefloat{figure}

 \vspace{0.25in}

\fontsize{14}{18}

\begin{document}
\title{Rare Speed-up in Automatic Theorem Proving Reveals Tradeoff Between
Computational Time and Information Value}
\author[1]{Santiago Hern\'{a}ndez-Orozco}
\author[2]{Francisco Hern\'{a}ndez-Quiroz}
\author[3,4]{Hector Zenil}
\author[5]{Wilfried Sieg}

\affil[1]{ \small Posgrado en Ciencias e Ingenier\'ia de la Computaci\'on, UNAM, Mexico.}
\affil[2]{ \small Departamento de Matem\'aticas, Facultad de Ciencias, UNAM, Mexico.}
\affil[3]{ \small Department of Computer Science, University of Oxford, UK.}
\affil[4]{ \small Algorithmic Nature Group, LABORES, Paris, France.}
\affil[5]{ \small Laboratory for Symbolic \& Educational Computing, Department of Philosophy, Carnegie Mellon University, USA.}

\date{}

\maketitle

\begin{abstract}
We show that strategies implemented in automatic theorem proving
involve an interesting tradeoff between execution speed, proving
speedup/computational time and usefulness of information. We
advance formal definitions for these concepts by way of a notion
of \emph{normality} related to an expected (optimal) theoretical
speedup when adding useful information (other theorems as axioms),
as compared with actual strategies that can be effectively and
efficiently implemented. We propose the existence of an
ineluctable tradeoff between this \textit{normality} and
computational time complexity. The argument quantifies the
usefulness of information in terms of (positive) speed-up. The
results disclose a kind of no-free-lunch scenario and a tradeoff
of a fundamental nature. The main theorem in this paper together
with the numerical experiment---undertaken using two different
automatic theorem provers ({\sc AProS} and {\sc Prover9}) on
random theorems of propositional logic---provide strong theoretical
and empirical arguments for the fact that finding new useful information
for solving a specific problem (theorem) is, in general, as hard
as the problem (theorem) itself.\\

\end{abstract}

\section{Introduction}

Speed-up in automatic theorem proving should not be regarded just as a quest for faster provers. To prove a theorem is an essentially difficult task and a very sensible question is whether there can be a shortcut to it by adding additional information. So the issue is what information can be added so that a proof is significantly shorter. Viewed from a very abstract point of view, of course, there is an easy (and trivial) answer: add the theorem itself and the proof becomes minimal! But adding all provable theorems to an axiomatic system renders it pointless.

On the other hand, a system under memory constraints that are not minimal (i.e. that can store a set of useful theorems apart from the axioms of a theory) but at the same time bounded by a \emph{realistic} measure, would be theoretically optimal if it stores only the most useful theorems for a certain task (that is, those that can make a proof shorter). A naive strategy would be to prove theorems incrementally and save them (until the memory is full). When a new theorem has to be proved the system can resort to previous theorems to see if a shorter proof can be obtained other than trying to prove the theorem from scratch. But how useful can this approach be? As we will show, it is not only naive and so expectedly not very useful, but it is also very rarely fruitful and in equal measure a waste of resources.

For this, we use a very basic but objective measure of speed-up, namely shortening the length of proofs. We sample random propositional calculus theorems and with the help of two automatic theorem provers ({\sc AProS} and {\sc Prover9}) we produced alternative proofs for them including as additional axioms other (random) valid formulas. Afterwards, we compare the length of the proofs to establish if a shorter one arose from the use of the new axioms. In order to select our samples, we restrict our analysis to a fragment of propositional logic, given by an upper bound on the number of atomic propositions and of logical connectives.

To find a proof for a theorem $\beta$ from a finite set
of formulas $t$ using an automatic theorem prover, one can always
start from $t$ and apply all transformation rules until
$\beta$ is generated, then pick out the sequence on the path from
$t$ to $\beta$ as proof. In practice, however, important
optimization strategies have to be implemented to avoid
exponential execution time even for the simplest of proofs. We
will explore how implementing these algorithmic strategies leads
to a compromise between various seminal complexity currencies,
shedding light on a possibly more general challenge in problem
solving, related to the usefulness of information and
computational time.

\theoremstyle{definition}
\newtheorem{speedupProp}[equation]{Definition}
\begin{speedupProp}\label{speedupProp}

Let $s$ be a finite set of formulas and let $s\vdash\beta$ stand
for the fact that $\beta$ is provable from $s$. $D(s \vdash\beta)$
will denote the length of the minimum proof for $s\vdash\beta$, as
given by the number of logical deduction steps used in the proof
(for example, the number of lines in a Fitch diagram).

Let $t$ and $t^\prime$ be two representations of the
class of equivalent theories $[t]=[t^\prime]$ (that is, $t$ and $t'$ are finite sets of axioms with an identical set of logical consequences, or \emph{theories}) such that
$|t|<|t^\prime|$. We define the \emph{speed-up delta} for
$t\vdash\beta$ between $t$ and $t^\prime$, or simply
\emph{speed-up}, as the function:
\begin{align*}
\delta_{\beta}(t,t^\prime) = 1 -
\frac{D(t^\prime\vdash\beta)}{D(t\vdash\beta)}.
\end{align*}
\end{speedupProp}

Let $P=\{\phi_i\}$ be a finite subset of the set of all valid
formulas in propositional logic and let $\beta \in P$ and $t
\subseteq P$. Now $\mathbb{A}$ will be the set of all logical
consequences (theorems) of $t$ in $P$. Finally, the set
$\mathbb{S}$ is the subset of $\mathbb{A}$ such that for a given
subset $A$ of $P$, with $\beta\notin{}A$, we have the following
property:
\begin{equation*}
\delta{}_{\beta}(t,t\cup{}A)>0.
\end{equation*}
Note that this property is equivalent to $D(t \vdash\beta) > D( t\cup{}A \vdash\beta)$

We ask after the relation and distribution between the size of the
set $\mathbb{S}$ with respect to the size of a finite subset of
$\mathbb{A}$ as a function of the set $A$. In other words, given a
random axiomatic system, if we strengthen it by adding a number of
theorems as axioms, what does the distribution of non-trivial
speed-ups look like? A set of valid propositional calculus
sentences and axiomatic systems is constructed given three
variables: $n$ represents the maximum depth of composition of
logical operators, $m$ states the maximum number of different
literals that can be present in a sentence, and $j$ determines the
number of axioms in an axiomatic system.

We will report not only that instances of non-trivial positive
speed-up are relatively rare, but that their number is considerably smaller than the number of instances of negative speed-up. In other words, strengthening a
random axiomatic system by adding theorems as axioms tends to
increase the length of proofs found by automatic theorem provers.
We believe that the behavior observed verifies the stated
condition and that its oblique distribution is strongly related to
the computational difficulty of finding a proof of minimum size
and deciding the usefulness of information as a whole.

\section{Methodology}

{\sc Prover9} is an automated theorem prover for First-order and
equational logic developed by William McCune~\cite{prover9}  at
the Argonne National Laboratory. {\sc Prover9} is the successor of
the Otter theorem prover. {\sc Prover9} is free and open source. {\sc AProS} (Automated Proof Search) is a theorem prover that aims
to find normal natural deduction proofs of theorems in propositional
and predicate logic~\cite{AProS}. It does so efficiently for
minimal, intuitionistic and classical versions of first-order logic.

We undertake an empirical exploration of the speed-up in
propositional logic, using {\sc AProS}
and {\sc Prover9} in order to approximate the proof complexity
over randomly generated sets of axiomatic systems and
propositions. We later propose a necessary but possibly insufficient
property (in definition \ref{normal}) for judging the adequacy
of the approximations so obtained.\\

To perform the exploration described, we have first to narrow the
search space: We denote the set of all propositions bounded by $n$
and $m$ by $P(n,m)$, the set of all theories in $P$ by
$\mathbb{T}$ ($\mathbb{T}=2^P$), while $\mathbb{T}(n,m)$ and
$\mathbb{A}(n,m)$ denote the sets of all theories and arguments
bounded by $n$ and $m$ respectively. Finally, we define the sets:
\begin{equation*}
\mathbb{T}(n,m,j) = \{t : t\in \mathbb{T}(n,m) \text{ and } |t|=j\}
\end{equation*}
\noindent{}and
\begin{equation*}
\mathbb{A}(m,n,j)=\{t\vdash{}\beta_l : t \in \mathbb{T}(n,m,j)
\text{ and } \beta_l \in P(n,m) \}.
\end{equation*}

\noindent{}The sets $P(n,m)$ are generated recursively as follows:
\begin{equation}\label{en}
P ( n, m ) = S \cup \text{ } \{ op(a, b) : op \in{} \{\iff,
\implies, \wedge, \vee \} \text{ and } a, b \in S\}\text{ },
\end{equation}
\noindent{}where $S = P ( n - 1, m ) \cup Neg ( P ( n - 1, m ) $,
$Neg(X) = \{\sim{}a | a\in{}X\}$ and $P(0,m)$ is the set that
consists of the first $m$ variables of $P$.\\

Note that, given an order to the set of Boolean operators, we can
define an enumeration of all the members of $P$ based on its
\emph{generation order}, an order which is consistent among all
the $P(n,m)$ subsets for a given $m$. The order defined is
equivalent to giving each $P(n,m)$ set an array structure on which
the first position correspond to the first element of $P(n,m)$ and
$f(n,m)$ to the last, where $f$ is defined as
\begin{equation*}
f(n,m)=|P(n,m)|.
\end{equation*}
We call this array the \emph{propositions array}. Given this
order, we can represent each theory comprising $\mathbb{T}(n,m)$
by a binary string of size $f(n,m)$ on which the $i$th bit is $1$
iff $\phi_i$ is in the theory. Following the previous idea, we can
efficiently represent the members of $\mathbb{T}(n,m,j)$ by an
array of $j$ integers, where each integer denotes the number of 0s
present between each 1. Hence we can represent the theories $t \in
\mathbb{T}(n,m,j)$ by integer arrays of the form $K =
[k_1,...,k_j]$. We call this the $j$-representation of the theory
and denote it by $t_K$.\\

\theoremstyle{definition}
\newtheorem{complSint}[equation]{Definition}
\begin{complSint}\label{complSint}
If $\phi\in{}P(n,m)$ and $\phi\not\in{}P(n-1,m)$, then $n$ is
known as the \emph{syntactic complexity} of $\phi{}$ or depth of
$\phi{}$; it is denoted by $s(\phi{})$.
\end{complSint}

\noindent{}Now let's consider the following function:
\begin{equation*}
gs (t_K) = \displaystyle\sum_{i=1}^j k_i.
\end{equation*}
\noindent{}The function $gs$ gives us the \emph{degree of
separation} between the first element in the array of propositions
([0,...,0]) and $t_K$. Moreover, there is an exponential number of
theories that share a value for the function. We call such a set,
defined by $[g] = \{t | gs(t) = g\}$, a \emph{separation
class} and $g$ the separation order of all the theories in the
class. Recall that the position of each proposition depends on its
order of generation. Hence the syntactic complexity of each theory
is set by its order of separation.

\theoremstyle{definition}
\newtheorem{complSintTeo}[equation]{Definition}
\begin{complSintTeo}\label{complSinTeo}
We define the \emph{syntactic complexity} of a theory $t$, denoted
as $s(t)$, as
\begin{equation*}
s(t) = max \{s(\phi{}): \phi\in{}t\},
\end{equation*}
\noindent{}where $s(\phi{})$ is the \emph{syntactic complexity} of
$\phi{}$.
\end{complSintTeo}
\noindent{}Note that $t\in{}[s(t)]$.\\

The size of the systems we are exploring grows by $\sim{}
2^\frac{(16 m)^{2^n}}{16}$, making an exhaustive exploration
intractable. The methodology used therefore consists in sampling
the space of valid propositional sentences. The propositions are
then used as theorem candidates against subsets of the formulae
used as theories.\\

In order to compute each sample set we build two sets: a sample
set of theories denoted by $T$, composed of $x$ number of theories
in $\mathbb{T}(n,m,j)$, and the sample set of prospective theorems
denoted by $O$, composed of $|O|$ numbers of propositions in
$P(n,m)$. Each set is randomly generated by, first, choosing a
random list of numbers of the respective lengths between $1$ and
$f(n,m)$. For the list $O$, each of these numbers represents the
prospective theorems sampled (for each theory). For the list $T$,
the numbers represent a separation class from which we choose a
theory by assigning random numbers to each of the parameters of
its $j$ representation, with the condition that their sum is the
value of the chosen class. The chosen lists are then rid of
inconsistent theories for $T$ and inconsistent propositions with
respect to the first element for the list
$O$.\\

Afterwards, we use the lists obtained to compute a sample set of
$\mathbb{A}(n,m,j)$. First, for each $t\in{}T$ we generate an
additional number of theories of the form $t\cup{}O_{j'}$, where
$O_{j'}$ is a prefix of the list $O$; we call $t=t\cup{}O_0{}=
t\cup{}\emptyset$ a \emph{base theory} and $t\cup{}O_{j'}$ a
\emph{derived theory}. Then, we pair all the theories generated
with each of the propositions of $o_{j}\in{}O$, called
\emph{objectives}, to form a list of cases of the form
$t\cup{}O_{j'}\vdash{}o_j$. Afterwards, we remove the unprovable
cases using by exhaustively exploration the corresponding
truth tables.\\

It is important to note that we are generating a significant number of
trivial cases when $j' = j$. Hence we expect at least close to
$\sim{}|O|\times{}|T|$ instances of positive speed-up, depending
on the number of unprovable cases.\\

Finally, we run an automatic theorem prover (ATP) and register the
length of each of the proofs obtained, storing the shortest ones.
However, we have no reason to believe that the use of an ATP
system would give us a good approximation to the sparsity of
proving speed-up. Therefore we define a speed-up function
relative to each ATP:

\theoremstyle{definition}
\newtheorem{speedupPropR}[equation]{Definition}
\begin{speedupPropR}\label{speedupPropR}
Let $A$ be an {\sc ATP} system, $t\vdash\beta$ a provable argument
for $A$, $t$ and $t^\prime$ two descriptions of $[t]$ such that
$|t|<|t^\prime|$. We define the \emph{speed-up delta relative to $A$} of
$t\vdash\beta$ between $t$ and $t^\prime$, or simply \emph{relative
speed-up}, as the function:
\begin{align*}
\delta_{{A(\beta)}}(t,t^\prime) = 1 -
\frac{D_{A(\beta)}(t^\prime\vdash\beta)}{D_{A(\beta)}(t\vdash\beta)}.
\end{align*}
\end{speedupPropR}
\noindent{}where $D_{A(\beta)}(t\vdash\beta)$ is the shortest
proof found by $A$ for the argument $t\vdash\beta$.

\theoremstyle{definition}
\newtheorem{invarianciaRelC}[equation]{Definition}
\begin{invarianciaRelC}\label{invarianciaRelC}
Let $L$ be a formal system, $A$ be an {\sc ATP} system for $L$,
$t\vdash{}\beta$ a provable argument for $A$, and $t^\prime$ a
description for $[t]$ such that $t\subset{}t^\prime$. We call the
function $\epsilon_t :\mathbb{N} \rightarrow{} \mathbb{R}$ a
\emph{bound} for $A$ as a function of $t$ if

\begin{equation*}
\delta_{A(\beta)}(t,t^\prime)\geq{}0 + \epsilon_t(|t^\prime|).
\end{equation*}
\end{invarianciaRelC}

Now, since we do not have enough information about the existent
sparsity of proving speed-up, we will define a necessary but
possibly insufficient condition needed for an
acceptable approximation:

\theoremstyle{definition}
\newtheorem{normal}[equation]{Definition}
\begin{normal}\label{normal}
We say that an argument $t\vdash\beta$ is \textit{trivial} if
$\beta \in t$. An {\sc ATP} system is \emph{possibly normal} for
$L$ if, for each $t$, $\epsilon_t(x)=0$, and if $t\vdash{}\beta$
is a non-trivial
argument, then $\delta_{A(\beta)}(t,t\cup\{\beta\})>0$.\\
\end{normal}

(In lemma \ref{primerProp} we will show that normality is possible, although our example is far from ideal.)

The mathematical structure used to store and analyze the results
obtained is called the \emph{speed-up matrix}. On this matrix each
entry has assigned the value

\begin{equation*}
\delta_{i,j}= \delta_{o_j}(t(i,j),t_i)
\end{equation*}
\noindent{}where
\begin{equation*}\label{matTi}
t(i,j)= \left\lbrace\begin{array}{l l} t_i & \text{ if $t_i$ is a base theory.}\\
t:D(t\vdash{}o_j)=\text{$\min$}_{D(t\vdash{}o_j)}\{t_k,..,t_{i-1}\} &\text{ otherwise.}\\
\end{array}
\right.
\end{equation*}

\noindent{}If $t_i \vdash{} o_j$ is not a provable case, then the
value of the entry is left undefined.\\

A natural simplification of this matrix is the \emph{incidence
speed-up matrix}, on which we simply assign different discrete
values to one of the following four cases: $\delta_{i,j} > 0$,
$\delta_{i,j}=0$, $\delta_{i,j}<0$ and $\delta_{i,j}$ is
undefined.\\

Note that by design we expect a prevalence in both matrices
of diagonal structures composed of cases of positive speed-up with
a period of $|O|$. These structures correspond to the
cases of trivial speed up included.\\

\section{Results}

To begin with, we performed more than 15 experiments using {\sc Prover9}.
The following table resumes a select number of results:\\

\begin{figure}[H]\label{tabla1}
\centering
\begin{tabular} {l|c|c|c|c|c}
\hline
\textbf{Exp. Num.} & \textbf{Cases} & $\delta > 0 $ & \textbf{Percentage} & $\delta < 0$ & \textbf{Ratio} \\
\hline
11 & 5400 & 606 & 11.2\% & 94 & 6.44\\
10 & 6381 & 704 & 11.01\% & 137 & 5.138 \\
7 & 4848 & 389 & 8.02\% & 231 & 1.683 \\
5 & 5454 & 426 & 7.81\% & 24 & 17.75 \\
3 & 11297 & 856 & 7.57\% & 70 & 12.228\\
\hline
\end{tabular}
\vskip 2ex
\caption{The results exhibit a varying percentage of
negative and positive speed-up instances. It is important to note
the presence of a significant number of negative speed-up instances
and the irregular distribution found among the samples.}
\end{figure}

\begin{figure}[H]\label{Prover91}
\begin{center}
\noindent{}\includegraphics[scale=2.5, trim={0 0 2 0},
clip=true]{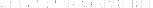} \vskip 0.2cm
\includegraphics[scale=2.5] {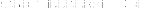}
\vskip 0.2cm
\includegraphics[scale=2.5] {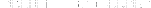}
\vskip 0.2cm
\includegraphics[scale=2.5] {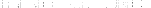}
\end{center}
\caption{A grayscale representation of the speed-up matrix obtained
for experiment number 11 using {\sc Prover9}. The columns correspond to each
theory generated and the rows to the theorems. For
visualization purposes, the matrix is divided
into four parts and only the instances of positive speed-up are colored,
the darker tones corresponding to higher speed-up values.}
\end{figure}

As the table shows, {\sc Prover9} does not exhibit normal behavior as
defined in \ref{normal}. Furthermore, as exemplified in Fig. \ref{Prover91},
the speed-up matrix does not present the expected
periodic diagonal speed-up instances.

Then, we present the \emph{Incidence Speed-up Matrix} obtained
from {\sc AProS} under four different conditions: the basic
experiment (the four logical connectives and classical deduction)
(\ref{1}); exclusion of disjunction as a connective (\ref{2}); the
basic experiment with intuitionistic logic (\ref{3}); intuitionistic
logic while restricting to the negative fragment as in \ref{2}
(\ref{4}). The same set of cases was used when possible, i.e.
\ref{1} with \ref{3} and \ref{2} with \ref{4}. Also included is
the matrix obtained for {\sc Prover9} during experiment \ref{1}.
It is important to note that we obtained no negative speed-up
values. The matrix values are represented using a color scale,
where the color white corresponds to no speed-up, blue to positive
speed-up and red to negative speed-up. Grey corresponds to
unprovable cases or cases in which the time limit
was reached.\\

The four experiments yield similar behavior: although {\sc AProS}
does show periodic diagonal structures it also exhibits a
significant presence of negative speed-up instances, which makes
the ATP otherwise than \emph{normal}.\\

It is important to note that a degree of clustering of negative
speed-up instances is expected, since each delta is computed from
the minimum proof length found for each previously derived theory
and current objective. It is arguable whether or not we are
overcounting negative speed-up instances.\\

Each \textit{speed-up matrix} is divided into four parts for
visualization purposes. Each of the entries' values is represented
using a four-color scale, where the color white corresponds to no
speed-up, blue to positive speed up and red to negative speed-up.
The columns correspond to each theory generated and the rows to
each of the theorems. Grey corresponds to unprovable cases or
cases where a time limit was reached.

\subsection{{\sc AProS} speed-up incidence matrix with classical
deduction and all four logical connectives} \label{1}

\noindent{}\textbf{Provable Cases:} 5763.\\ \textbf{Positive}
$\delta$: 632, \textbf{percentage:} 10.97\%.\\
\textbf{Negative} $\delta$: 564, \textbf{percentage:} 9.78\%.\\

\begin{figure}[H]\label{AprosClas}
\begin{center}
\noindent{}\includegraphics[trim={0 0 538.2 0}, clip=true,
scale = 2.5]{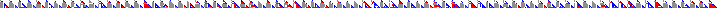}\\
\vskip 0.2cm
\includegraphics[trim={179 0 359.2 0}, clip=true, scale =
2.5]{Apros2NC.png}\\
\vskip 0.2cm
\includegraphics[trim={358.2 0 180.2 0}, clip=true, scale =
2.5]{Apros2NC.png}\\
\vskip 0.2cm
\includegraphics[trim={538.2 0 0 0}, clip=true, scale =
2.5]{Apros2NC.png}\\
\vskip 0.2cm
\end{center}
\caption{A color scale representation of the incidence speed-up
matrix obtained for experiment \ref{1} (classical deduction and
all four logical connectives) using {\sc AProS}. The periodic
diagonal structures that correspond to the trivial speed-up
instances are evident in this figure, but it also manifests a
significant presence of negative speed-up instances, which means
that the {\sc AProS} is not a \emph{normal} ATP
system.}\label{Apros1}
\end{figure}

\subsection{{\sc AProS} speed-up incidence matrix with intuitionistic
deduction and all four logical connectives:}\label{3}

\noindent{}\textbf{Provable Cases:} 5680.\\ \textbf{Positive}
$\delta$: 646, \textbf{percentage:} 11.37.\%\\
\textbf{Negative} $\delta$: 537, \textbf{percentage:} 9.45\%.\\

\begin{figure}[H]\label{AprosInt}
\begin{center}
\noindent{}\includegraphics[trim={0 0 532.2 0}, clip=true,
scale = 2.5]{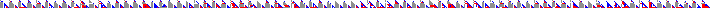}\\
\vskip 0.2cm
\includegraphics[trim={178.2 0 354.2 0}, clip=true, scale =
2.5]{Apros22NC.png}\\
\vskip 0.2cm
\includegraphics[trim={356.2 0 176.2 0}, clip=true, scale =
2.5]{Apros22NC.png}\\
\vskip 0.2cm
\includegraphics[trim={534.2 0 0 0}, clip=true, scale =
2.5]{Apros22NC.png}\\
\vskip 0.2cm

\caption{A color scale representation of the incidence speed-up
matrix compiled for experiment \ref{3} (intuitionistic deduction and
all four logical connectives) obtained from the same set of
arguments used to compile the figure \ref{Apros1}. The behavior is
very similar to behavior observed in experiment \ref{1}, along
with the significant presence of negative speed-up. We detected a
small increase in the number of positive speed-up instances and a
negligible decrease in negative speed-up cases.}\label{Apros2}
\end{center}
\end{figure}

\vskip 2ex

\subsection{{\sc AProS} speed-up incidence matrix with classical
deduction and without disjunction}\label{2}

\noindent{}\textbf{Provable Cases:} 6680.\\ \textbf{Positive}
$\delta$: 899, \textbf{percentage:} 13.46 \%.\\
\textbf{Negative} $\delta$: 484, \textbf{percentage:} 7.246\%.\\

\begin{figure}[H]
\begin{center}
\noindent{}\includegraphics[trim={0 0 501 0}, clip=true,
scale = 2.5]{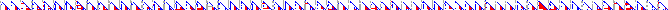}\\
\vskip 0.2cm
\includegraphics[trim={167 0 334.2 0}, clip=true, scale =
2.5]{Apros2NoConjC.png}\\
\vskip 0.2cm
\includegraphics[trim={334.2 0 167.2 0}, clip=true, scale =
2.5]{Apros2NoConjC.png}\\
\vskip 0.2cm
\includegraphics[trim={501.2 0 0 0}, clip=true, scale =
2.5]{Apros2NoConjC.png}\\

\caption{A visual representation of the incidence speed-up matrix
compiled for the experiment \ref{2} (classical deduction without
disjunction) divided into four parts for visualization purposes.
It is important to note that the set of arguments employed for the
previous experiments is incompatible with the parameters
established for this case. Hence a new random set had to be
generated. From the image we can see that the speed-up
distribution does not differ notably from previous experiments,
aside from the significantly lower incidence of undemonstrable
arguments.}\label{Apros3}

\end{center}
\end{figure}

\vskip 2ex

\subsection{AProS speed-up incidence matrix with intuitionistic
deduction and without disjunction} \label{4}

\noindent{}\textbf{Provable Cases:} 6660.\\ \textbf{Positive}
$\delta$: 862, \textbf{percentage:} 12.94\%.\\
\textbf{Negative} $\delta$: 587, \textbf{percentage:} 8.81\%.\\

\begin{figure}[H]
\begin{center}
\noindent{}\includegraphics[trim={0 0 499 0}, clip=true,
scale = 2.5]{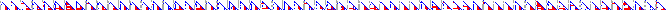}\\
\vskip 0.2cm
\includegraphics[trim={167 0 332.2 0}, clip=true, scale =
2.5]{Apros22NoConjC.png}\\
\vskip 0.2cm
\includegraphics[trim={334 0 165.2 0}, clip=true, scale =
2.5]{Apros22NoConjC.png}\\
\vskip 0.2cm
\includegraphics[trim={501 0 0 0}, clip=true, scale =
2.5]{Apros22NoConjC.png}\\

\caption{A visual representation of the incidence speed-up matrix
generated for experiment \ref{4} (intuitionistic deduction
without disjunction) obtained from the same set of arguments used
to compile the figure \ref{Apros3}. We can see that the behavior
is very similar to that observed in experiment \ref{2}, along with
the significant presence of negative speed-up. We detected a
negligible decrease in the number of positive speed-up instances
and a small increase in negative speed-up cases.}\label{Apros4}
\end{center}
\end{figure}

\vskip 2ex

\subsection{{\sc Prover9} incidence speed-up matrix without disjunction:}\label{5}

\noindent{}\textbf{Provable Cases:} 6680.\\ \textbf{Positive}
$\delta$: 312, \textbf{percentage:} 4.67\%.\\
\textbf{Negative} $\delta$: 0, \textbf{percentage:} 0\%.\\

\begin{figure}[H]
\begin{center}
\noindent{}\includegraphics[trim={0 0 499 0}, clip=true,
scale = 2.5]{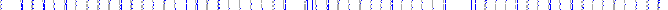}\\
\vskip 0.2cm \includegraphics[trim={167 0 332.2 0}, clip=true,
scale =
2.5]{NoConj9IncC.png}\\
\vskip 0.2cm \includegraphics[trim={334 0 165.2 0}, clip=true,
scale =
2.5]{NoConj9IncC.png}\\
\vskip 0.2cm \includegraphics[trim={501 0 0 0}, clip=true, scale =
2.5]{NoConj9IncC.png}\hspace{5pt}\\
\end{center}
\caption{A color scale representation of the incidence speed-up
matrix obtained for experiment \ref{1} (classical deduction with
all four logical connectives) using {\sc Prover9}. The image is
divided into four parts for visualization purposes. As with figure
\ref{Prover91} the absence of the predicted diagonal structures is
conspicuous, but of greater importance is the total absence of
instances of negative speed-up.}\label{Prover92}
\end{figure}

\subsection{Observations and Conclusions}

The main objective of this project was to undertake an empirical exploration of the prevalence
and distribution of instances of positive speed-up found within the propositional calculus.
In particular, two deduction systems for propositional logic were explored: natural deduction and binary
resolution, each of which was approximated by two automated proving systems, {\sc AProS} and {\sc Prover9}.
A necessary (but not sufficient) condition was proposed in order to decide the adequacy of
these approximations (Def. \ref{normal}).\\

\noindent{}Given the speed-up matrices obtained, it is evident
that neither {\sc AProS} nor {\sc Prover9} conforms to the
normality condition defined :\\

\noindent{}{\sc Prover9} cannot detect trivial cases with regularity;
instead of the expected periodic diagonal patterns induced by
the presence of instances of trivial speed-up, we find a number of
vertical clusters of speed-up instances without a discernible
regular distribution. This behavior is incompatible with the
second condition of normality. We also found a non-negligible
number of negative speed-up instances when a disjunction is
included in the list of logical connectives. The presence of the
disjunction seems to have little effect on the
distribution of instances of positive speed-up.\\

\noindent{}{\sc AProS} shows an important number of instances of negative speed-up
(slow-down). While {\sc AProS} does not have problems detecting cases of trivial
speed-up, the number of instances of negative speed-up is greater than in {\sc Prover9}.
Furthermore, the presence and distribution of these instances
is not significantly affected by the presence or absence of the disjunction,
nor by the alternation between intuitionistic and classical deduction.\\

We consider the observed behaviors as evidence of the computational
complexity that the proposed condition of normality entails:
\emph{discerning the usefulness of new information is intrinsically
computationally complex}.
We formalize this in the following statements:

\theoremstyle{plain}
\newtheorem{primerProp}[equation]{Lemma}
\begin{primerProp}\label{primerProp}
There is a normal prover.

\begin{proof}
Given the argument $t\vdash{}\beta$, a brute force algorithm that,
starting from the list $t$, simply searches for $\beta$
while building the (infinite) tree of all
possible logical derivations in a breadth-first search fashion,
will always find the proof of minimal length in the form of the selected branch.
\end{proof}
\end{primerProp}

\noindent{}The expected computational time of this algorithm is of the order
$O(q(D(t\vdash{}\beta))^{D(t\vdash{}\beta)})$, where $q$ is a polynomial that depends
on the number and structure of the derivation rules used.

\theoremstyle{plain}
\newtheorem{segundaConjetura}[equation]{Theorem}
\begin{segundaConjetura}\label{teorema}
Given a non-trivial argument $t\vdash{}\beta$ and a non-empty set
$t^\prime\subset{}P$ such that $t\subset{}t^\prime$, deciding if
$\delta_{\beta}(t,t^\prime)>0$ is $NP$-Hard.
\begin{proof}
Given the results found in \cite{alekhnovich} we can say that, if
$P\neq{}NP$, there is no polynomial time algorithm that can find
$D(\emptyset\vdash{}\beta)$ (if it is polynomial with respect to
$|\beta|$). However, if we can decide $\delta_{\beta}(t,t\cup
A)>0$ in polynomial time then we can find $D(\emptyset\vdash{}\beta)$
in polynomial time:\\

The algorithm we propose iterates the answer to
$\delta_{\beta}(\emptyset{}, \{\phi\})>0$ on each possible $\phi$
derivation from the list of chosen formulas $D$ ($D$ starts with
the list of axioms). The number of derived formulas is polynomial
with respect to $|D|$. Each of the positive instances is added to
a list $l$. The formula in $l$ that minimizes the proof length can
be found in $log(|l|)O(|l|)$ by pairwise comparison using the
result of $\delta_{\beta}(\{\phi_i{}\}, \{\phi_i,\phi_j{}\})>0$
and $\delta_{\beta}(\{\phi_j{}\}, \{\phi_i,\phi_j{}\})>0$ for each
$\phi_i,\phi_j\in l$. Note that if both values are TRUE, then both
formulas must be part of the smallest proof that contains any of
the propositions, so we can choose to add just one formula and the
other one will eventually be added on to the following iterations
if it is part of the smallest demonstration. We add the selected
expression to the list $D$.\\

Following the stated procedure, we will eventually reach a trivial
argument, finishing the demonstration of the tautology in the form
of list $D$ (the list contains a succession of formulas derived in
order from $t$). Note that we consult a polynomial time algorithm
a polynomial number of times over the size of a list that grows up
to a polynomial size with respect to the input. Hence the
algorithm finds the smallest proof in polynomial time.
\end{proof}
\end{segundaConjetura}

\noindent{}Given the demonstrated difficulty of the problem, we make the following conjecture:

\theoremstyle{plain}
\newtheorem{nuevaConjetura}[equation]{Conjecture}
\begin{nuevaConjetura}\label{nuevaConjetura}
There is no normal proving algorithm significantly faster than the brute force
algorithm described in \ref{primerProp}.
\end{nuevaConjetura}

In other words, we are proposing the existence of an ineluctable
tradeoff between normality and execution time of an
automated proving algorithm. The conjecture \ref{nuevaConjetura}
also implies that, for {\sc AProS} and {\sc Prover9}, the function
$\epsilon_t$ is of exponential order for each $t$.\\

Finally, in the context of an argument $t\vdash{}\beta$, we can
say that a set $A$ \emph{contains useful information} if it
induces (positive) speed-up \footnote{This means that
$\delta_{\beta}(t,t \cup A)>0$)} and that the information is
\emph{useless} otherwise. With the experiment and the theorem
\ref{teorema} we have presented empirical and theoretical
arguments as to why discerning the usefulness of new
information for solving a specific problem is as hard as the
problem itself.\\

As for the differences found in the speed-up matrices for {\sc
Prover9} and {\sc AProS}, we believe that these emerge mostly due
to an initial syntactic analysis performed by {\sc AProS} that
allows it to detect trivial cases; the exception being when
removing disjunction which results in no slow-down instances,
although this change doesn't seem to affect the positive speed-up
distribution in a significant way. The conjecture
\ref{nuevaConjetura} suggests that, for {\sc AProS} and {\sc
Prover9}, the function $\epsilon_t$ is of exponential order for
almost all $t$'s, and that both are within a linear constant
between them, else we should be able to find a shortcut to
normality.\\

Whenever the result presented in figure \ref{Prover92} is a
counterexample to this statement and the conjecture
\ref{nuevaConjetura} is an open question.  We could argue that
simplifying the formulas by removing number of logical connectives
is doing the prover's job. And, if we do restrict our space to
simpler (yet complete) set of formulas, a stronger normality
condition should be able to be defined as:

\theoremstyle{definition}
\newtheorem{normalHard}[equation]{Definition}
\begin{normalHard}\label{normalHard}
A system is \emph{normal} for $L$ if there exist a polynomial
time algorithm that calls {\sc ATP} as an oracle for deciding
$\delta_{\beta}(t,t^\prime)>0$, with $t$, $t'$ and $\beta$ as in
theorem \ref{teorema}.
\end{normalHard}

\end{document}